\documentclass[conference, anonymous=false]{IEEEtran}
\IEEEoverridecommandlockouts
\usepackage{cite}
\usepackage{amsmath,amssymb,amsfonts}
\usepackage{algorithmic}
\usepackage{graphicx}
\usepackage{textcomp}
\usepackage{xcolor}

\usepackage{caption}
\usepackage{subfigure}

\usepackage{comment}
\usepackage{marvosym} 
\usepackage{hyperref}
\hypersetup{
hypertex=true,
colorlinks=true,
linkcolor=purple,
anchorcolor=purple,
citecolor=purple
}

\def\LOGO#1{\textsc{#1}}

\def\BibTeX{{\rm B\kern-.05em{\sc i\kern-.025em b}\kern-.08em
    T\kern-.1667em\lower.7ex\hbox{E}\kern-.125emX}}
\begin{document}

\title{LUDA: Boost LSM Key Value Store Compactions with GPUs\\
}

\author{
{
Peng Xu,  
Jiguang Wan\textsuperscript{\Letter}, 
Ping Huang\textsuperscript{\ddag}, 
Xiaogang Yang, 
Chenlei Tang, 
Fei Wu, 
and Changsheng Xie}\\
\IEEEauthorblockA{\textsuperscript{\Letter} Corresponding author\\
\textit{Wuhan National Laboratory for Optoelectronics, and School of Computer Science and Technology}\\
\textit{Huazhong University of Science and Technology, Wuhan, Hubei, P. R. China}\\
\textit{\textsuperscript{\ddag} Department of Computer and Information Sciences, Temple University, Philadelphia, Pennsylvania, USA}\\
\{jgwan, wufei, cs\_xie\}@hust.edu.cn
}
}


\maketitle

\begin{abstract}


Log-Structured-Merge (LSM) tree-based key value stores are facing critical challenges of fully leveraging the dramatic performance improvements of the underlying storage devices, which makes the compaction operations of LSM key value stores become CPU-bound, and slow compactions significantly degrade key value store performance.
To address this issue, we propose \LOGO{Luda}, an \underline{L}SM key value store with C\underline{UDA}, which uses a GPU to accelerate compaction operations of LSM key value stores.
How to efficiently parallelize compaction procedures as well as accommodate the optimal performance contract of the GPU architecture challenge \LOGO{Luda}.
Specifically, \LOGO{Luda} overcomes these challenges by exploiting the data independence between compaction procedures and 
using cooperative sort mechanism and judicious data movements.
Running on a commodity GPU under different levels of CPU overhead, evaluation results show that \LOGO{Luda} provides up to 2x higher throughput and 2x data processing speed, and achieves more stable 99th percentile latencies than LevelDB and RocksDB.

\end{abstract}


%
%

\section{Introduction}

%

\begin{figure}[htbp]
  \centering
  \includegraphics[scale=0.6]{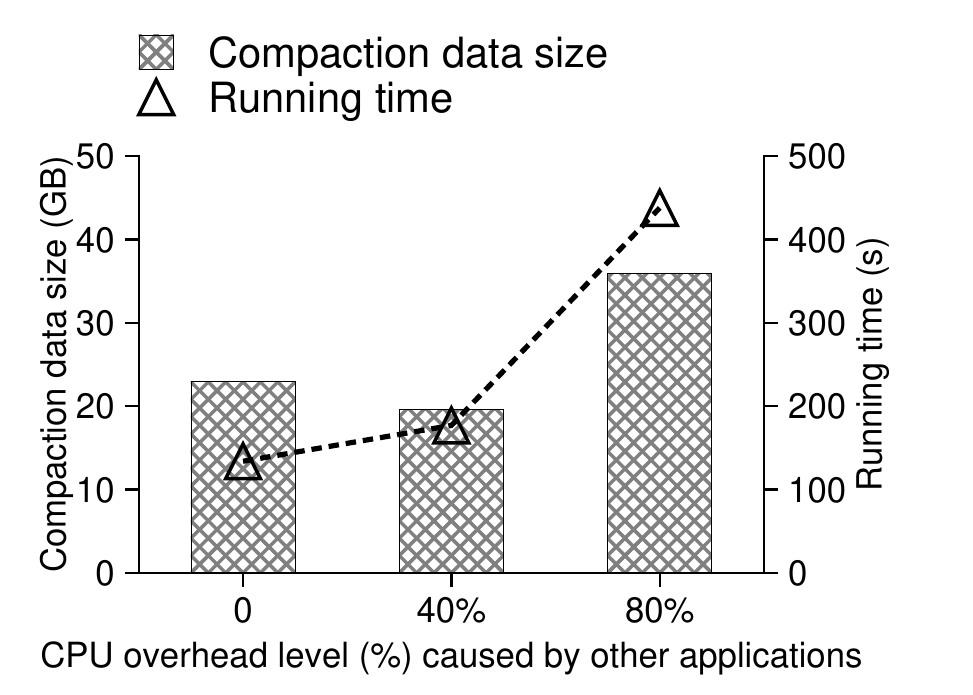}
\caption{RocksDB performance under different CPU overhead.}
\label{fig-motivation-cmp}
\end{figure}

Key-Value (KV) stores empower a wide range of modern internet services in large scale data centers, such as e-commerce \cite{SOSP07Dynamo} and web indexing \cite{URL19RocksDB}.
Among KV stores equipping with diverse elaborated indices like B+ trees and Log-Structured Merge (LSM) trees, LSM KV stores are widely adopted in write-heavy workload environments \cite{ATC19SILK} in the age of slow storage devices.
LSM KV stores avoid in-place-update via batching KV pairs to leveled data layouts in  storage devices, and later digest KV pairs between levels to guarantee read performance as well as to reclaim storage space (namely compaction).
Propogate dispersed KV pairs, which causes significant performance variances, is less I/O-bound and becomes increasingly CPU-bound  \cite{SOSP19KVell, IPDPS14LSM} with the rapid performance improvements of storage devices, such as NVMe Solid State Drives (SSDs).
Take the performance of a widely used LSM KV store RocksDB \cite{URL19RocksDB}  in Figure \ref{fig-motivation-cmp} as an example, we use different levels of CPU overhead to simulate CPU resource contention from other applications and use an Optane SSD as the high-performance storage device.
When the overall CPU overhead is 80\%, the running time  of this 4-thread RocksDB takes 3x longer time while the size processed by compaction is about 1.5x more.
The significant running time increment shows that the CPU becomes the bottleneck of compactions.


This transition of the compaction bottleneck has led to research attentions on techniques that alleviate the CPU overhead, like 
non-sorted KV designs \cite{SOSP19KVell, ATC18HashKV}, 
in-drive or remote node KV pair processing schemes \cite{ICPP19TStore, VLDB15Compaction}. 
However, it is necessary to do compactions for LSM KV stores.
First, it dramatically reduces the cost-effectiveness if not recyclying the storage space occupied by stale key values.
Second, even though high-end SSDs provide continuous and stable performance at the device level, it is difficult to guarantee that the underlying persistent layer, such as file systems, maintains good performance with fragments.
Third, as DRAM memory performance will degrade at high space utilization \cite{FAST14LogDRAM}, it is reasonable to assume that high-performance SSDs are unable to prevent performance drop with few free spaces \cite{FAST12SFS}.
When the memory is scarce and key values for accessed in a random manner, KV stores not conducting compaction shows significant performance degradation \cite{SOSP19KVell}.
Conducting compactions is necessary to ensure stable performance in the long run.

Offloading compaction jobs from the CPU provides a good remedy for compaction induced performance drops \cite{SIGMOD19X-Engine, ICPP19TStore} since compactions are CPU-bound.
The compaction is the only way to put all inserted key values into SSTs as well as making room for new SSTs at the lower levels.
Slow compactions decrease the write performance or even disrupt.
However, spending more CPU resources to speed up compactions suffers from two concerns.
First, one of the important metrics for most existing server schedulers is the high CPU utilization, which makes it difficult to cost-effectively ensure adequate CPU time for compactions timely.
Second, the marginal revenue of allocating more CPU cores for compactions at a time diminishes because of NUMA effects.
Since compactions are not on the read path, it is feasible to process compactions separately to improve write performance.
The key for compactions is to generate new SSTs as fast as possible, but generating new SSTs has to calculate checksums for all data blocks as well as iterating all keys to get bloom filters and miscellaneous blocks. 
These operations are compute-intensive.
Leveraging SSD in-drive processors to generate SSTs shares the computation overhead of the CPU and reduces data transfers, however, the limited power budget and relatively slow computation speed make it easy to surpass their potentials.
X-Engine \cite{FAST20FPGA-LSM} uses specialized FPGAs for compactions, which can provide considerable computational power.
Inspired by the search for more computational power, we turn to GPUs to boost LSM KV compactions, which are widely used in computation power hungry scenarios.
On the one hand, GPUs are more development-friendly when compared to specializing FPGAs.
On the other hand, continuous improvements and emerging applications make GPUs a promising heterogeneous computation supplement to CPUs \cite{ASPLOS15GPUfs, SC18DRAGON}.

In this paper, we propose \LOGO{Luda}, LSM KV stores with CUDA, to overcome the compaction computation-bound bottleneck.
However, existing key value store compaction procedures are optimized for CPUs, while the GPU has a different architecture and contracts of optimal performance.
How to take advantage of the GPU's abundant computational power by efficiently parallelizing compaction operations with different optimal memory access requirements is challenging.
By developing \LOGO{Luda}, we present the main contributions as follows:

\begin{enumerate}
\item To the best of our knowledge, we are the first to use the GPU to boost LSM KV store compactions.
\item We exploit the parallelism potentials of compaction procedures and accommodate the GPU memory hierarchy with minimum data movements.
 \item We prototype \LOGO{Luda} with CUDA based on LevelDB by modifying about 1K lines of code and adding 2.7K lines of CUDA code and validate the design of \LOGO{Luda} with the widely used YCSB benchmark.
\end{enumerate}

The rest of the paper is organized as follows.
We first describe the background of LSM KV stores and related architectural features of GPUs in Section \ref{sec-background}.
Section \ref{sec-design} explains the design of \LOGO{Luda}, 
and Section \ref{sec-evaluation} evaluates and analyzes the prototype performance.
We briefly describe related work in Section \ref{sec-related-work}, and finally conclude in Section \ref{sec-conclusion}.


\begin{figure}[htbp]
  \centering
  \includegraphics[width=3.34in]{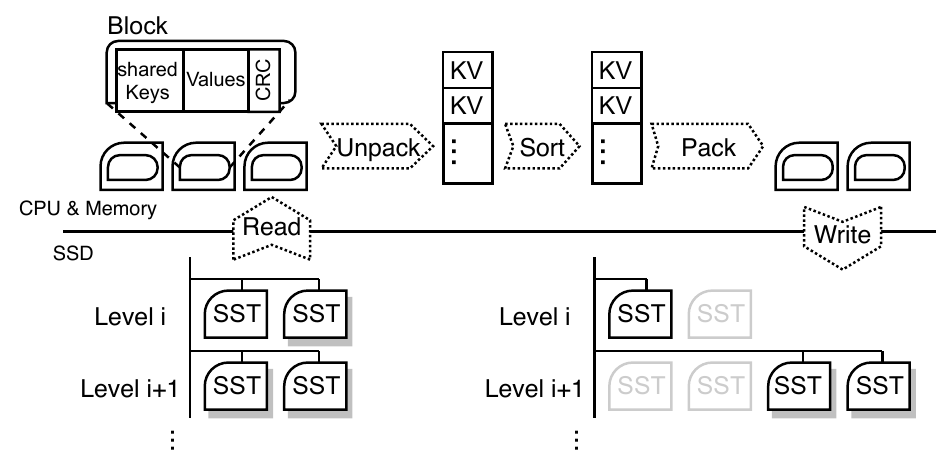}
\caption{LSM-tree overview and compactions.}
\label{fig-lsm}
\end{figure}

\begin{figure}[htbp]
  \centering
  \includegraphics[width=2.5in]{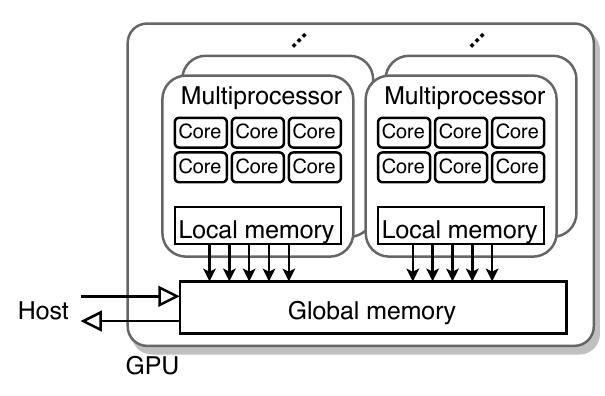}
\caption{Overview of the simplified GPU architecture.}
\label{fig-gpu}
\end{figure}

\section{Background}
\label{sec-background}

\subsection{LSM-tree Compaction}
\label{subsec-lsm-and-compact}

Figure \ref{fig-lsm} outlines the LSM tree and its compaction procedures with the example of LevelDB.
Key value pairs are squeezed in blocks to reduce storage space usages.
In a data block, the keys of key value pairs are stored in the shared keys part, which eliminates their redundant prefixes.
The values are stored in the value part.
Finally, a checksum (typically CRC32) is used for the shared keys and value part.
One or more data blocks along with indices of their key value pairs and a bloom filter block, which is used for checking the existence of a key in this SST, form a sorted-string table (SST), and LevelDB writes to storage devices in the unit of an SST.

LevelDB tiers SSTs with different size limitations for each level, and moves key value pairs from level i to level i+1.
Newly generated SSTs by user inserts are written to level 0.
When the size of level i reaches its quota, LevelDB selects SSTs from level i and those SSTs in level i+1 with overlapped key ranges, then deletes stale key value pairs as well as sorting them.
After that, new SSTs are generated and written to level i+1.
Finally, it reclaims old SSTs in level i.
This compaction mechanism makes level i the stage area for level i+1, and guarantees there is free space available for new SSTs from level i-1.
Note that, until successfully attaching new SSTs to LevelDB, key values in old SSTs of level i are valid and used for retrieving when the compaction is in processing.


Specifically, the compaction procedures include the following phases:

\textit{Phase 1 Unpack:}
After reading selected SSTs, LevelDB has to restore key value pairs from data blocks.
First, LevelDB recalculates the checksum of each data block for data integrity.
Second, LevelDB iterates the shared keys and restore key value pairs, and now these keys are available for sorting.

\textit{Phase 2 (Delete and) Sort:}
Since restored key value pairs from selected SSTs have key overlaps, LevelDB merges and sorts all key value pairs and drops those marked with delete.
Then the key-sorted key value pairs are ready to build new SSTs.

\textit{Phase 3 Pack:}
LevelDB groups several key value pairs into a data block  according to the defined data block size.
Then, it begins to calculate the shared keys in each data block.
After that, it calculates the checksum for each block.
Finally, LevelDB groups several data blocks into an SST, generates indices for them, and calculates the bloom filter blocks for this SST.

\subsection{GPU Basic}
\label{subsec-gpu-abc}

Figure \ref{fig-gpu} gives a simplified  memory hierarchy of a GPGPU.
Each core in the GPU is a data processing unit, and a GPU has many cores that provide powerful computational capability.
Cores are organized in the unit of a Stream Multiprocess (SM), and each SM provides fast local memories for these cores.
All cores in the same SM handles different data objects but running the same instruction simultaneously, and the local memory access the larger and slower global memory in a batch way.
The GPU exchanges data with the host in the global memory, and both the bandwidth and latency of the global memory is very high.
Besides, existing data prefetching features in the GPU is not able to efficiently handle sophisticated control flows and data caches between the SM local memory and the global memory.

CUDA \cite{URL19CUDA} provides APIs to program for general purposes with NVIDIA GPUs.
CUDA batches a number of threads (typically 32) as a scheduling unit (warp) and dispatches warps to SMs, and all threads in the same warp manage their own and share data with other threads through the SM local memory.
One or more warps form a thread block, and one or more thread blocks form a CUDA kernel.


\section{Design}
\label{sec-design}

In this section, we first explain \LOGO{Luda}'s overall workflow, then discuss design challenges and concerns, and finally discuss the details on how these concerns are embedded in \LOGO{Luda}'s design.

\begin{figure*}[!htbp]
\centering
\includegraphics[scale=0.93]{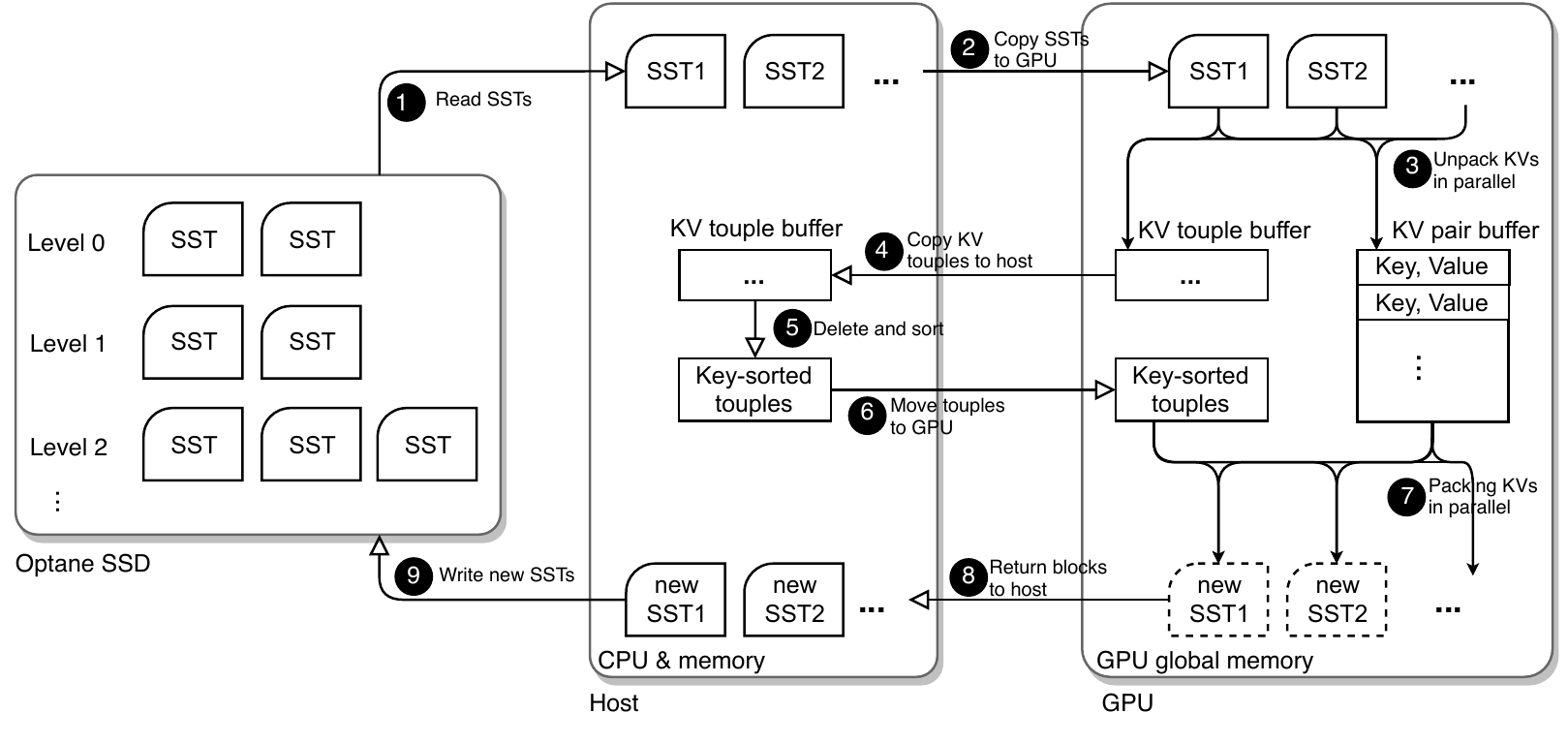}
\caption{Overview of \LOGO{Luda}.}\label{fig-LUDA-overview}
\end{figure*}

\subsection{\LOGO{Luda} Overview}
\label{subsec-LUDA-overview}

%

Figure \ref{fig-LUDA-overview} shows the overview workflow of bringing the GPU into compaction operations with \LOGO{Luda}.
Currently, the CPU plays the role of coordinator between the high-performance Optane SSD and the GPU, and SSTs are stored in the Optane SSD, and the GPU completes the majority of compaction operations  for selected SSTs.
\LOGO{Luda} executes following compaction procedures for each compaction job:

\begin{enumerate}
\item Initiate the compaction procedures and read selected SSTs from the Optane SSD to memory in parallel.
\item Copy SSTs from memory to the GPU.
\item Unpack key value pairs from SSTs, and gather all pairs in the KV pair buffer.
Meanwhile, \LOGO{Luda} also generates key value tuples in the KV tuple buffer, and use tuples to sort and locate key value pairs.
\item Transfer key value tuples from the KV tuple buffer to the CPU memory.
\item Use the CPU to delete stale key value tuples and sort them.
\item Copy key-sorted tuples to the GPU.
\item Use key-sorted key value tuples to pack key value pairs from the KV pair buffer to fix-sized data blocks and bloom filter blocks (we use the bloom filter block as the representative of metadata blocks in an SST unless otherwise noted).
\item Return prepared data and bloom filter blocks to the host, and the CPU writes these new SSTs to the Optane SSD.
\end{enumerate}

As explained in the steps above, the CPU is free from unpacking and packing key value pairs to SSTs, and only involves operations in sorting the key value tuples, which greatly alleviates the computational overhead on the CPU.
However, still, the CPU plays as the transfer station between the Optane SSD and the GPU, and we discuss this effect on \LOGO{Luda} designs later in Section \ref{subsec-LUDA-movingCost}.
To the right of Figure \ref{fig-LUDA-overview}, \LOGO{Luda} utilizes the GPU to handle almost all compaction operations.
\LOGO{Luda} boosts up to 2x performance with the help of powerful computation capabilities of the GPU when the CPU is under heavy workloads.
However, the memory hierarchy of the GPU differs from the CPU memory hierarchy that multiple cores do the same job on different data, hence requiring careful compaction operation map-reduce strategies and crafting memory accesses accordingly.

\subsection{\LOGO{Luda} Design Principle}
\label{subsec-LUDA-principle}

Shifting compaction jobs to GPU is not without challenges, and this subsection describes the major challenges of \LOGO{Luda} designs.
\LOGO{Luda} exploits GPU computing capability to reduce compaction computation overheads.
Further, \LOGO{Luda} utilizes GPU's many cores and high bandwidth to parallelize compaction operations and reduce response time.

\textit{Challenge 1}: How to parallelize compaction operations to the best effort?
Even though compaction operations are mainly composed of three phases as presented earlier, pipelining them \cite{IPDPS14LSM} in coarse granularities fails to exploit full computational potentials of the GPU.
\LOGO{Luda}  further refines operations of these phases and parallelize them with different CUDA kernels.
How to compose efficient CUDA kernels for better parallelism to free the  computation capability of GPU?

\textit{Challenge 2}: How to minimize data movement cost?
The overhead of moving data inside the GPU or from the host to the GPU is non-trivial.
If we do not take this overhead into \LOGO{Luda} design considerations, the benefit of parallelizing compaction operations will be compromised.

\begin{figure*}[htbp]
  \centering
  \subfigure[Unpack key value pairs from SSTs in parallel.]{
    \begin{minipage}{2.8in}
      \centering
      \includegraphics[width=2.7in]{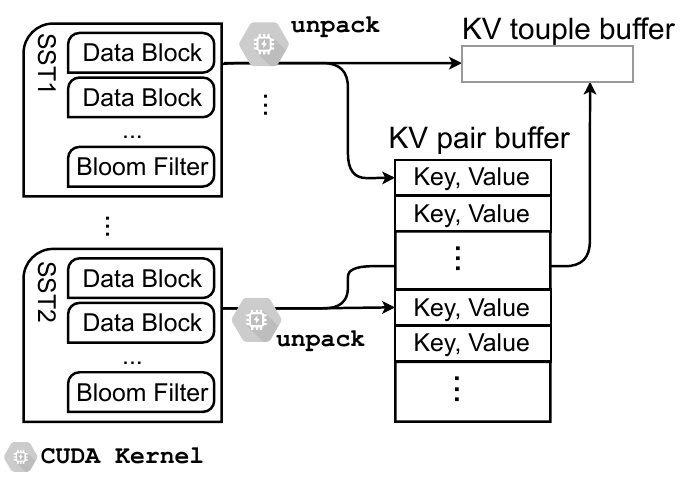}
    \end{minipage}
    \label{fig-unpack}
  }
  \subfigure[Packing key value pairs to SSTs in parallel.]{
    \begin{minipage}{4in}
      \centering
      \includegraphics[width=4in]{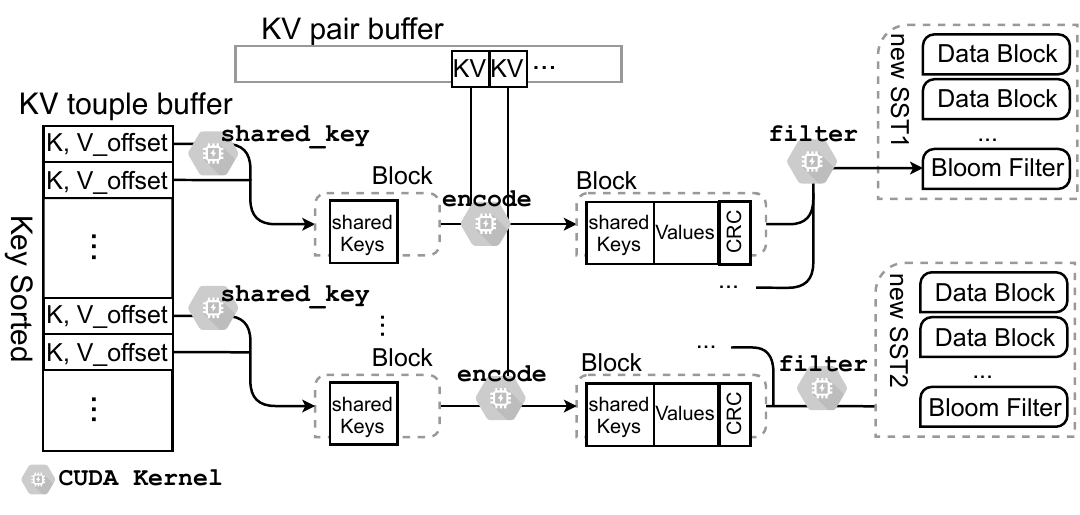}
    \end{minipage}
    \label{fig-package}
  }
\caption{CUDA kernels of \LOGO{Luda} and workflows for different compaction operations.}
\label{fig-packageUnpack}
\end{figure*}

\subsection{Parallelizing Compaction Work}
\label{subsec-LUDA-parallel}


To parallelize compaction operations, we first discuss data dependencies in the three phases of compaction.
In phase 1 unpacking key value pairs from SSTs, \LOGO{Luda} has to check the checksum of each data block to ensure the data integrity of squeezed key value pairs in it, then restores key value pairs with their original keys.
The inputs for unpacking operations are mainly individual SST data blocks, and there are no data dependencies among data blocks.
Operations in phase 3 reverse the job done in phase 1, however metadata blocks in SSTs depend on all data blocks.
Exploiting parallelism for compaction operations of unpacking and packing key value pairs is viable, but need to be designed differently.
In phase 2 sorting key value pairs, \LOGO{Luda} will merge and sort all the restored key value pairs from phase 1 in their key order, and data dependences of overlapped key ranges from different SSTs cause inevitable data exchanges, which will bring considerable data movement overheads inside the GPU (discussed in Section \ref{subsec-LUDA-movingCost}).

For unpacking operations, \LOGO{Luda} uses an \verb|unpack| CUDA kernel for each SST as illustrated in Figure \ref{fig-unpack}, and the hexagon represents a CUDA kernel.
In an \verb|unpack| kernel, \LOGO{Luda} checks the checksum of each data block and restores key value pairs in it in parallel, and puts restored key value pairs of this SST into the KV pair buffer.
\LOGO{Luda} gathers all restored key value pairs in this buffer, and generates tuple \verb|<K,V_offset>| for key value pairs and uses tuples to sort them to reduce data movements in the GPU.

In Figure \ref{fig-package}, \LOGO{Luda} uses three different CUDA kernels \verb|shared_key|, \verb|encode|, and \verb|filter| to improve the compaction operation parallelism for packing operations.
Each \verb|shared_key| kernel uses sorted KV tuples to calculate the shared keys of key value pairs for each data block.
Then \verb|encode| kernels fill data blocks with corresponding values and calculate the checksum (here is CRC32) for each data block.
Finally, \verb|filter| kernels generate bloom filters blocks for each SST.
 \verb|shared_key| kernels enable the batch access to the KV tuple buffer and parallelize the building of the  shared\_key part of different data blocks.
There are no data exchanges among \verb|encode| kernels.
When building the bloom filter block with kernels \verb|filter|, \LOGO{Luda} initiates transfering prepared data blocks to the host.
With the above three CUDA kernels based on refined operations of different phases, \LOGO{Luda} pipelines these parallelized compaction operations.

%

\subsection{Reducing Data Access and Movement Cost}
\label{subsec-LUDA-movingCost}


%

%

Shifting compaction jobs from the CPU to a GPU introduces challenges of data movements inside and outside the GPU.

\paragraph{Data moving between the CPU and GPU} 
One challenge is how to efficiently move data from the CPU to the GPU.
Ideally, there is no need to move compaction involved SSTs to the CPU, because 
the LSM decouples the path of reading inserted key values and existing SSTs are able to provide valid key values before they are replaced by compaction generated SSTs.
Freeing storage space from invalid key values in selected SSTs and generating new SSTs are off the critical indexing path of finding requested key values.
Thus directly moving old and newly generated SSTs between the storage device and the GPU is able to reduce both the memory and computation overhead of the CPU. 
The newly release GPUDirect Storage \cite{URL19GPUDirectStorage} and AMD SSG \cite{URL16AMDSSG} technologies enable the direct data path between storages and GPUs.
However, we are unable to access these featured GPUs for now, and \LOGO{Luda} is orthogonal to these optimizations. 
We plan to embed this feature into \LOGO{Luda} in the future.

\begin{figure}[htbp]
  \centering
  \subfigure[Copy SSTs to the GPU.]{
    \begin{minipage}{2.6in}
      \centering
      \includegraphics[width=2.6in]{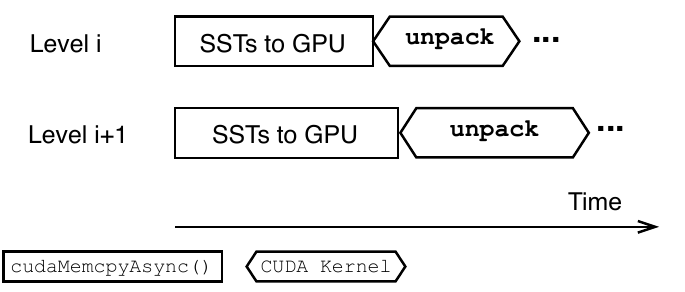}
    \end{minipage}
    \label{fig-toGPUpipeline}
  }
  \subfigure[Copy new SST blocks to the CPU.]{
    \begin{minipage}{2.6in}
      \centering
      \includegraphics[width=2.6in]{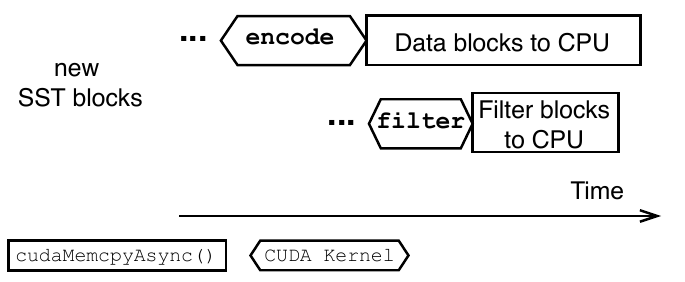}
    \end{minipage}
    \label{fig-toCPUpipeline}
  }
\caption{Transferring data between the CPU and GPU in parallel.}
\label{fig-packageUnpack}
\end{figure}

Currently, \LOGO{Luda} moves SSTs between the Optane SSD and the GPU via the CPU memory.
To boost the data transfer, \LOGO{Luda} asynchronizely moves SSTs of the lower and upper levels independently.
As illustrated in Figure \ref{fig-toGPUpipeline}, \LOGO{Luda} initiates two threads to copy SSTs of the lower level $Li$ and the upper-level $L_{i+1}$ in parallel, and starts \verb|unpack| kernels for them upon their arrival at GPU.
When returning compaction results of new SSTs to the CPU, there is more space to copy blocks in parallel.
\LOGO{Luda} uses different CUDA kernels to generate new SST data and filter blocks, and preparing different blocks takes different time, and \LOGO{Luda} begins the transfer of prepared data blocks from the GPU to the CPU.
In Figure \ref{fig-toCPUpipeline}, when an \verb|encode| kernel completes building data blocks, \LOGO{Luda} triggers the movements of them to the GPU.
Meanwhile, \LOGO{Luda} also continues building filter blocks of these data blocks,  copies the built filter block to the CPU, and composes new SSTs in the host.
As the size of filter blocks are significantly smaller than the data blocks', the overhead of transferring filter blocks are overlapped.
By decomposing and overlapping data transfers, \LOGO{Luda} mitigates the time overhead of data movements between the CPU and GPU.

\paragraph{Data moving within the GPU} 
How to efficiently move data inside the GPU is the other challenge.
Even though providing significant high throughput between outer devices, GPUs require to coalesce data movements inside to deliever better performance.
However, this requirement challenges the parallelizing compaction operations for the following reasons.
The first reason is that the size of the high-performance local memory of an SM is about tens of kilobytes, and it is not large enough to host all SSTs because the size of an SST is usually several megabytes.
Therefore, \LOGO{Luda} will inevitably access the slower GPU global memory to fetch demanded key values to sort them.
The second reason is that operations of phase 2 merging and sorting key values that cause scattered memory accesses.
When there are overlapped key ranges among SSTs, which is usually the fact, \LOGO{Luda} has to copy demanded key values from different SSTs of different locations to gather them in a buffer, and this means scattered small size memory accesses to the GPU global memory, which will consume significant time.
We observe that sorting \verb|<K,V_offset>| tuples or \verb|<Key, Value>| pairs with a GTX 1080Ti takes up to milliseconds even for only thousands of them.
This will significantly compromise the benefits of exploiting compaction parallelism.

Taking the above requirements and limitations into consideration, \LOGO{Luda} introduces  the following design techniques to achieve efficient data movements.


\begin{itemize}
\item \textit{Lightweight Sort Mechanism}: 
Considering that values are only used for checksums and do not affect phase 2 merge sort, \LOGO{Luda} decouples values from their keys when unpacking them from old SST data blocks with CUDA kernel \verb|unpack|.
Kernel \verb|unpack| puts restored key value pairs to the KV pair buffer as shown in Figure \ref{fig-unpack}.
Meanwhile, it also generates \verb|<K, V_offset>| tuples to represent key value pairs and stores them in the KV tuple buffer.
\verb|K| is the restored key and \verb|V_offset| locates this key value pair in the KV pair buffer.
\LOGO{Luda} sorts these key value tuples instead of key value pairs and greatly reduces the amount of unnecessary data movements.
However, sorting these small size tuples still bottlenecks the overall compaction operations in the GPU as explained earlier, therefore \LOGO{Luda} employs a the cooperative sort mechanism.
\item \textit{Cooperative Sort Mechanism}: 
To our best knowledge, we do not find an efficient CUDA library to sort \verb|<K, V_offset>| tuples and plan to improve this in the future.
Currently, \LOGO{Luda} cooperates with CPU for sorting them.
After building \verb|<K, V_offset>| tuples, \LOGO{Luda} transfers them to the CPU asynchronously and fetches them back to the GPU for composing SST blocks.
Since the size of tuples is about tens of kilobytes for megabytes of SSTs, this asynchronous data transfer is acceptable, and this method avoids the slow data movements inside the GPU.
More importantly, we later find out that this cooperative sort does not introduce significant latency overhead in Section \ref{subsec-lat-perf}.
\item \textit{Lazy Value Movement}: 
After getting through three compaction phases and four different CUDA kernels, \LOGO{Luda} moves values with the following two times.
First, unpack key value pairs from the original SSTs and put them to the KV pair buffer.
Second, pack key value pairs from the KV pair buffer moves values to new data blocks with the help of key-sorted KV tuple buffer.
These two necessary movements for values enable the parallelism of unpacking and packing operations and also minimize the overhead of data movements inside the GPU. 
\end{itemize}


\section{Evaluation}
\label{sec-evaluation}


Our evaluation of \LOGO{Luda} aims to demonstrate the design insights in speeding up compaction procedures and improving the performance of LSM key value stores.
We answer the following questions using \LOGO{Luda}:

\begin{enumerate}
\item What are the benefits of introducing the GPU to replace the CPU for compaction operations?
\item Does the computation power of a GPU improve the compaction performance?
\item How effective is \LOGO{Luda}'s compaction parallelism in reducing request latency?
\end{enumerate}


\subsection{Configuration and Methodology}
\label{subsec-configure}


\paragraph{Server Setting}
We conduct evaluations on one server running Linux Ubuntu 16.04 with kernel 4.15.
The server uses the Supermicro X10DAI motherboard with one 4-core (8-thread, and the full CPU utilization is 800\%) Intel E5-1620 v4, 64 GB DRAM, one NVIDIA GTX 1080Ti GPU, and one 280 GB Intel Optane 900p SSD.

\paragraph{Software Setting}
We prototype \LOGO{Luda} with CUDA (version 10.1) based on LevelDB (version 1.22) \cite{URL19LevelDB} by modifying about 1K lines of code and adding 2.7K lines of CUDA code, and we are ready to open source it\footnotemark.
We implement bloom filter calculations in \LOGO{Luda}, but leave the snappy compression and plan for the future work.
We believe that if we enable the snappy compression, evaluation results will strengthen the advantage of \LOGO{Luda}.
\footnotetext{\href{https://github.com}{https://github.com/anonymous}}

In all tests, we use LevelDB (version 1.22) as the baseline and RocksDB (version 6.4.6) as the well-optimized LSM-tree key value store.
All key value stores are configured with default parameters and 10 bits for bloom filters, except that RocksDB is configured with four threads to represent high CPU demanding key value stores.

We evaluate \LOGO{Luda}, LevelDB, and RocksDB using YCSB cloud benchmark \cite{SoCC10BechmarkYCSB}.
We set key size 16 B for all key-values and only vary the value size (from 128 B to 1 KB).
The size of both an memtable and a SST is 4 MB, and the size of a data block is 4 KB unless explicitly claimed.
In all experiments, we use the YCSB A workload, which has a 0.5-0.5 update-read ratio.
We first load 10 million key value pairs into each key value store, which has a total database size of about 5 GB.
Then we play another 10 million key value pairs on all stores.
Finally, we set the value size of 256 B and use 0.1 billion key value pairs (total 50 GB) as the number for YCSB to stress all key value stores and check their realtime latency and CPU/GPU usages.

To simulate scenarios that other applications compete for or exhaust the CPU, we use stress-ng \cite{URL19stress-ng} to generate different CPU overhead levels (40\% and 80\%, and the full CPU overhead level setting is 100\%).

\begin{figure*}[!htbp]
\centering
\includegraphics[scale=0.55]{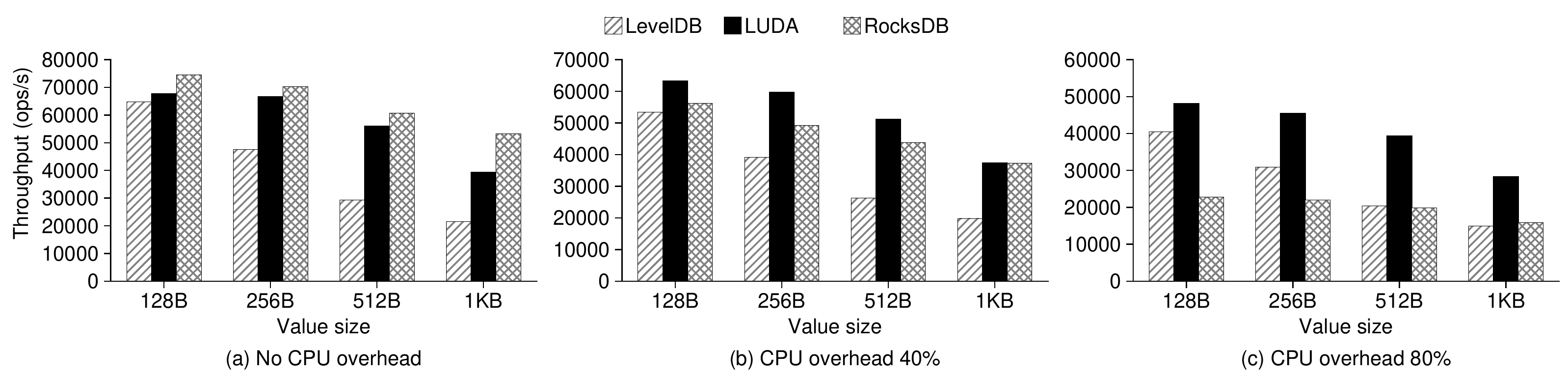}
\caption{Throughput (ops/s) under different CPU overhead.}\label{fig-iops-sst4}
\end{figure*}

\begin{figure*}[!htbp]
\centering
\includegraphics[scale=0.55]{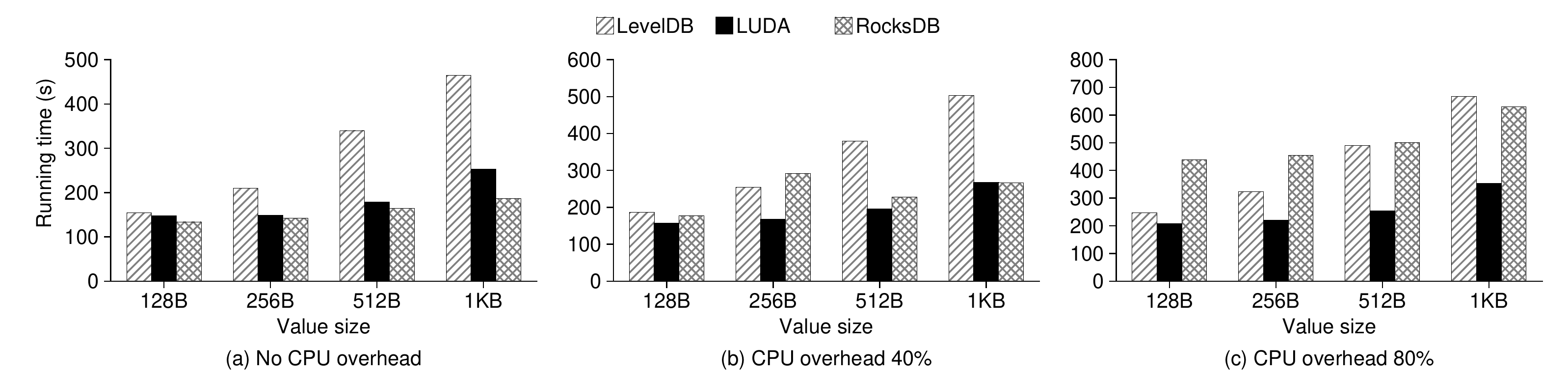}
\caption{The execution time (s) under different CPU overhead.}\label{fig-time-sst4}
\end{figure*}

\begin{figure*}[!htbp]
\centering
\includegraphics[scale=0.55]{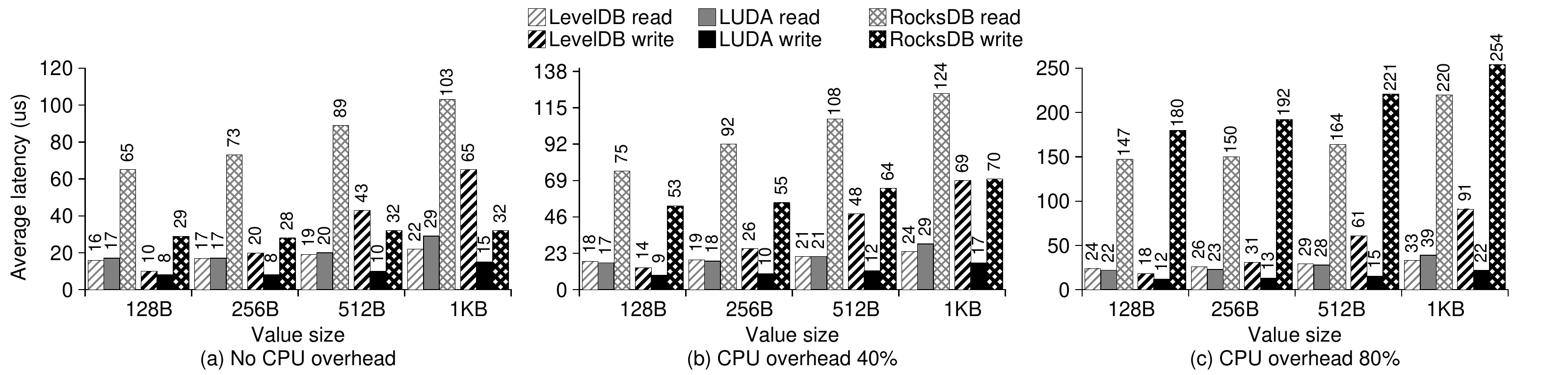}
\caption{The average read and write latency (us) under different CPU overhead.}
\label{fig-lat-mean}
\end{figure*}

\begin{figure*}[!htbp]
\centering
\includegraphics[scale=0.55]{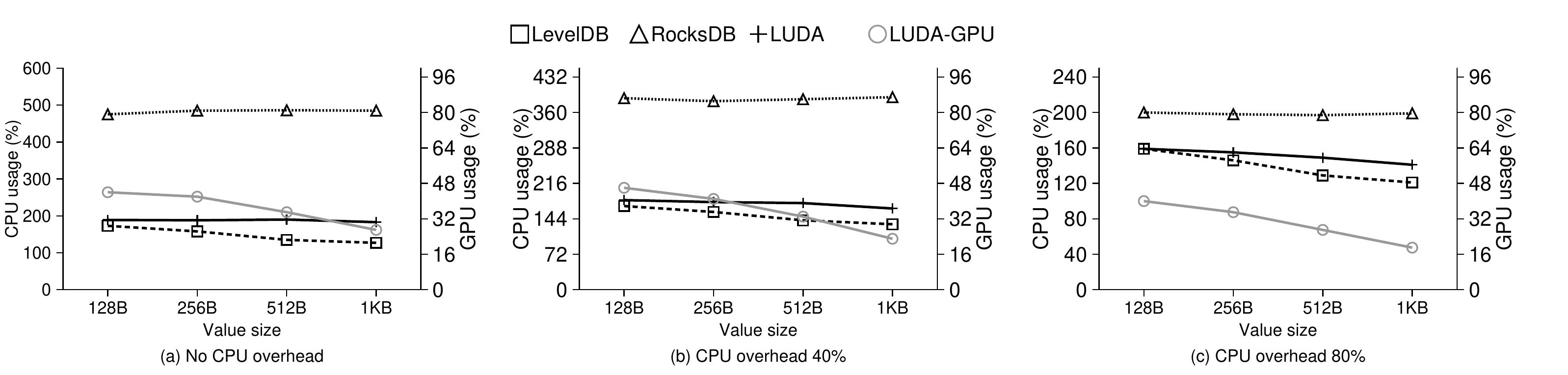}
\caption{The CPU/GPU usages under different CPU overhead.}\label{fig-cpu-sst4}
\end{figure*}


\subsection{Overall Performance}
\label{subsec-overall-perf}

The throughput (ops/second), running time, and request latency show the effectiveness of using the computation capability of the GPU to replace the CPU computation in compaction operations.
We set different CPU stress levels and compare the throughput and execution time of all key value stores.

Figure \ref{fig-iops-sst4} shows the throughput results under different CPU overhead levels (0, 40\%, and 80\%).
With the growth of CPU overhead, all key value stores show lower throughputs.
There are two reasons: 
(1) Half of all requests are read, and these reads are slowed down with fewer CPU time, and \LOGO{Luda} does not touch the read path; and (2) CPU still involves compaction operations more or less, and when other applications take too many CPU resources, these key value stores are unable to handle compaction requests.
As \LOGO{Luda} depends on the CPU for sorting key value tuples, this also contributes to the reduced throughput of \LOGO{Luda}.
When the CPU overhead is 80\% with value size 1KB, RocksDB only provides about 30\% of the throughput when there is no CPU overhead.
\LOGO{Luda} and LevelDB maintain about 70\% of the throughput without CPU overhead, but \LOGO{Luda} shows about 2x  the throughput when compared to LevelDB and RocksDB with only 20\% available CPU resources.
The higher throughput of \LOGO{Luda} shows the effectiveness of migrating computational operations from the CPU to GPU, and also proves that GPU boosts the compaction operations.

Figure \ref{fig-time-sst4} gives the running time to complete a 5 GB data size with different CPU overhead.
When the CPU is free from distractions, RocksDB has the shortest running time with all value sizes, and \LOGO{Luda} follows RocksDB, and LevelDB is the slowest one.
With the increase of value sizes, all key value stores take a longer time.
LevelDB nearly doubles its running time, and RocksDB takes about 40\% more time.
When the CPU overhead is 80\%, all key value stores have longer running time compared to their results no CPU overhead.
At the value size 1 KB,  the running time of RocksDB grows about 2x, and LevelDB spends extra 203 seconds, and \LOGO{Luda} uses extra 100 seconds.
Although \LOGO{Luda} takes more time when there are fewer CPU resources, it is clear that \LOGO{Luda} is more robust to the CPU resource variance and outperforms others at all value sizes.

\begin{figure*}[!htbp]
\centering
\includegraphics[scale=0.55]{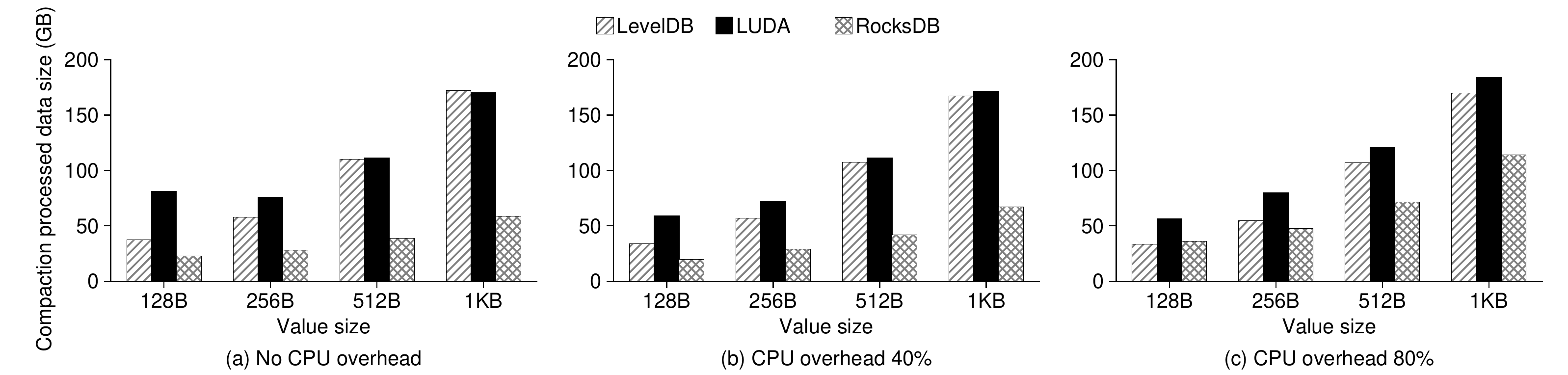}
\caption{The compaction processed data size under different CPU overhead.}\label{fig-cmpSize-sst4}
\end{figure*}

\begin{figure*}[!htbp]
\centering
\includegraphics[scale=0.55]{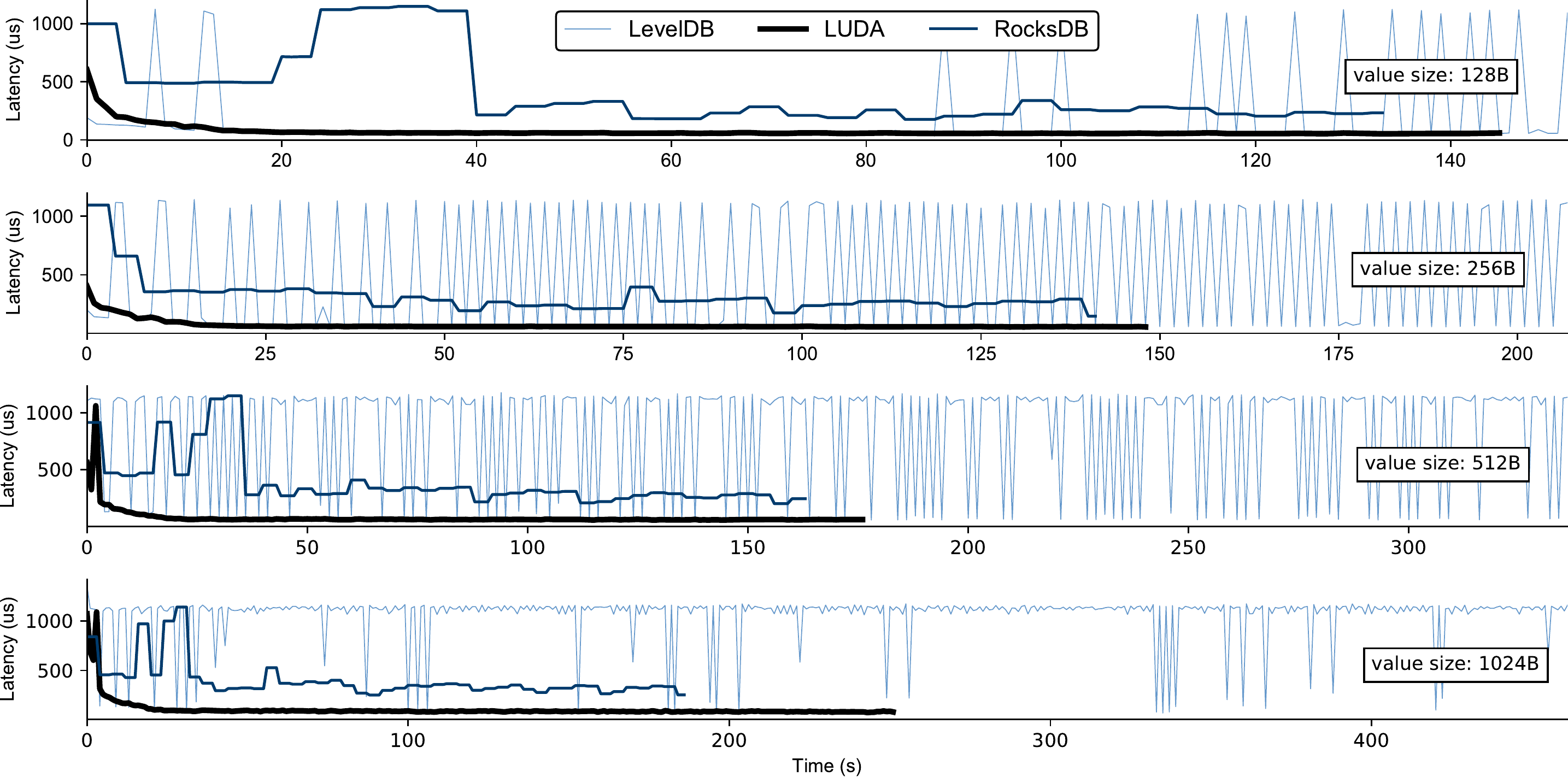}
\caption{99th realtime latency with different value sizes without CPU overhead.}\label{fig-realtime-lat-99th}
\end{figure*}

Figure \ref{fig-lat-mean} shows the average latency for read and write requests under different CPU overheads.
When there is no CPU overhead, average latencies of all key value stores increase along with the growth of value sizes.
When the value size is 1 KB, the average write latency of LevelDB increases by 55 us, and \LOGO{Luda} and RocksDB's average write latency have moderate growth.
When the CPU overhead is 80\%,  both LevelDB and RocksDB have longer average read and write latency than \LOGO{Luda}, and the average write latency of \LOGO{Luda} increases by only 7 us.
As for all average read latencies with different CPU overhead, RocksDB shows the higher read latency than \LOGO{Luda} and LevelDB.
We infer that request queues in RocksDB are responsible for it even though RocksDB uses four threads.
Since RocksDB uses more threads than \LOGO{Luda} and LevelDB, RocksDB average latencies are greatly affected by the CPU overhead.
Another observation is that \LOGO{Luda} almost suppresses the average write latency across different CPU overhead, which means that \LOGO{Luda} efficiently frees the CPU from compaction operations.

Figure \ref{fig-cpu-sst4} gives the CPU usage of all key value stores and the GPU usage of \LOGO{Luda}.
With larger value sizes, CPU usages and GPU usages of \LOGO{Luda} decrease.
This is because it takes more time for large value movements and the increased size of compaction involved data (in Figure \ref{fig-cmpSize-sst4}).
Larger values reduce the overhead of calculating bloom filters, therefore \LOGO{Luda} GPU usages decrease.
When the CPU overhead is heavy, the four-thread RocksDB almost saturates the remaining CPU resources.
The CPU usage of \LOGO{Luda} and LevelDB, as well as the GPU usage of \LOGO{Luda} marginally decreases.
However, \LOGO{Luda} consumes more CPU resources than LevelDB.
The reason for this moderate higher CPU usage of \LOGO{Luda} is due to the frequent data movements between the CPU and GPU for the larger size of compaction data.

\begin{figure*}[!htbp]
\centering
\includegraphics[scale=0.55]{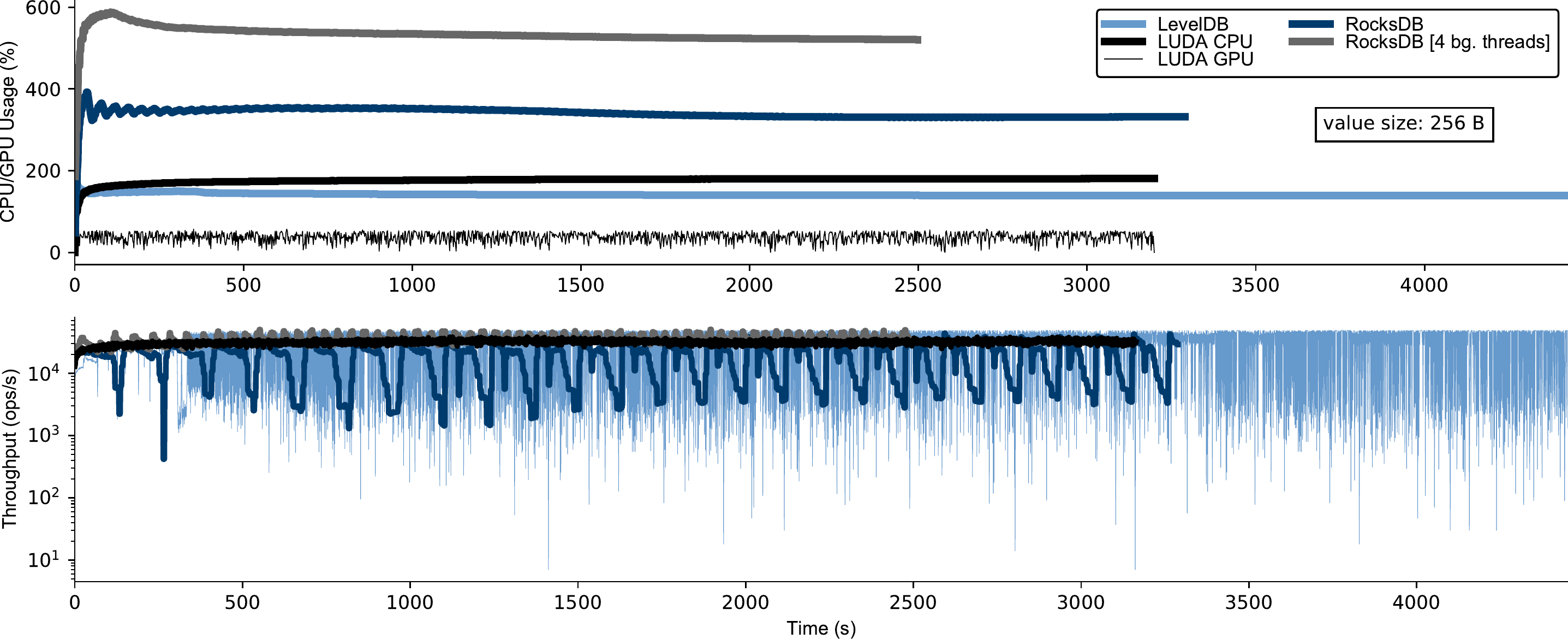}
\caption{Realtime CPU/GPU usage and throughput (ops/s) with 0.1 billion 256B-value key values and no CPU overhead.}\label{fig-cpu-longA}
\end{figure*}

\subsection{Compaction Speed}
\label{subsec-compact-perf}

We now use the size of compaction processed data by all key value stores to evaluate their compaction speed.
Both the data size of compaction induced read and write are included in Figure \ref{fig-cmpSize-sst4}, and all of them have the 0.5-0.5 ratio for the read size and write size.
It's clear that the CPU overhead has negligible effects on the compaction data size. \LOGO{Luda} and LevelDB involved much more data in compactions for all sizes of value, and \LOGO{Luda} handles the larges size of data.
We infer that the column family \cite{URL19RocksDB} in RocksDB helps reduce the overlapped range for higher-level SSTs.
As for the large \LOGO{Luda} compaction data size, we figure out that this is caused by our prototype simplification.
Currently, \LOGO{Luda} triggers the compaction procedures when Level 0 is full and does not dump the full immutable when conducting compactions in the GPU, so \LOGO{Luda} misses this chance of involving more SSTs.
As a consequence, the overlapped key range for the next compaction will be larger than other key value stores, hence causing more compaction data.
We plan to address this disadvantage of our prototype in the future.
However, the above results also prove that \LOGO{Luda} processes compactions much faster than LevelDB and RocksDB.
When the value size is 128 B, the size of data processed by \LOGO{Luda} is almost 2x the size of LevelDB, while the running time of \LOGO{Luda} is even shorter than the LevelDB's in Figure \ref{fig-lat-mean}.
When the value size is 1 KB, \LOGO{Luda} handles the comparative amount of data compared to LevelDB with 46\% less running time.
Results in Figures \ref{fig-cmpSize-sst4} and \ref{fig-lat-mean} show that, \LOGO{Luda} compaction speed is 1x faster than the speed of LevelDB and 2x faster than the speed of RocksDB.

\subsection{Latency Variance}
\label{subsec-lat-perf}

The latency variance is another important metric for compaction performance.
Figure \ref{fig-realtime-lat-99th} shows the 99th percentile latency without CPU overhead for different value sizes.
At the beginning of the running, \LOGO{Luda} shows a relatively high and short-time latency spike, this is because moving data to the GPU and initiating the parallel compaction in GPU takes time.
Finally, \LOGO{Luda} presents the lowest and smooth latency for all value sizes.
Without boosting compactions by a GPU, LevelDB shows significant latency variances when the value size gets larger and finally gets the highest 99th percentile latency.
RocksDB uses four threads and approximately suppresses the latency variance.
We further test the robustness of \LOGO{Luda} with the 0.1 million dataset size (about 50 GB) for about an hour as shown in Figure \ref{fig-cpu-longA}.
In this experiment, we increase the max background job of RocksDB to 4 to boost RocksDB compactions.
The CPU and GPU usage of all key value stores plotted on the top half of Figure \ref{fig-cpu-longA}, and plots stop when key value stores complete all requests.
As there is no extra CPU overhead, all key value stores show steady CPU usages.
\LOGO{Luda} consumes slightly more CPU resources than LevelDB, and the GPU usage of \LOGO{Luda} varies as called by compaction jobs from time to time.
RocksDB with 4 background jobs uses the most CPU resource.
The fewer compaction data , as well as more CPU resources, enables this 4-background-job RocksDB to take the shortest running time.
RocksDB with fewer background jobs presents visible throughput fluctuations.
\LOGO{Luda} provides stable and relatively high throughput when compared to the 4-background-job RocksDB.
\LOGO{Luda}'s stable and low 99th latency in Figures \ref{fig-realtime-lat-99th} and the high throughput in Figure \ref{fig-cpu-longA} confirm that handling compaction data quickly is helpful to stable performance.

\section{Related Work}
\label{sec-related-work}

Many pieces of research provide optimizations on the compaction of LSM-tree based key value stores.


\textit{Boost compaction operation:}
Computation is not the bottleneck of compactions when key value stores using slow storages, and Pipelined Compaction Procedure \cite{IPDPS14LSM} uses CPU pipelines to parallelize computation and I/O operations for faster compaction.
However, pipelining compactions only with CPU becomes CPU-bound when using fast storage devices like nowaday SSDs, let alone high-performance Optane SSDs.
Both SILK \cite{ATC19SILK} and Luo \cite{arXiv19LsmStability} analyze the I/O scheduling effects on compactions and use I/O schedulers to avoid write interference between user writes and compaction writes.
\LOGO{Luda} introduces the GPU into key value stores, and moves almost all compaction jobs to the GPU, and provides a solution to fast compactions for key value stores with high-performance storage devices.

\textit{Offload compaction:}
KVSSD \cite{DATE18KVSSD} embeds a key value store in the SSD Flash Translation Layer, and utilizes the FTL to complete all compaction jobs inside the SSD.
TStore \cite{ICPP19TStore, arXiv18Co-KV} dynamically partitions compaction tasks and conducts compactions by both SSDs and the host in parallel.
Doing compaction in SSDs reduces the size of transfer data between the host and storage, however the wimpy onboard processor and limited power budget challenge this kind of design as concerned in Summarizer \cite{MICRO17Summarizer}, which makes trade-offs between near SSD processing and the data transfers to the host.
X-Engine \cite{FAST20FPGA-LSM} uses FPGAs for compactions to reduce CPU consumption, and the FPGA is less developing-friendly.
Ahmad and Kemme \cite{VLDB15Compaction} leverage remote compute node to do compaction, which will be affected by networking.
\LOGO{Luda} firstly uses the GPU to boost compaction processings, as GPUs have better computation power than FPGAs or SSD onboard processors.

\textit{Reduce compaction data size:}
TRIAD \cite{ATC17TRIAD} reduces the compaction data size by keeping hot key values in the memtable and unifying the WAL and Level 0.
PebblesDB \cite{SOSP17PebblesDB} uses guards to reduce overlapped ranges between levels and improves the compaction process.
SifrDB \cite{SoCC18SifrDB} combines the multi-stage tree and the multi-stage forest and reduces write amplification.
WiscKey \cite{FAST16WiscKey} separates keys and values and reduces value movements for compactions. 
HashKV \cite{ATC18HashKV} optimizes WiscKey garbage collections for value logs by segmenting value logs and placing values in segments with hash.
dCompaction \cite{Springer17dCompaction} defers the processing for data blocks to prevent key values from being compacted too frequently. 
LDC \cite{ICDE19LDCCompaction}, LWC-store \cite{MSST17Wan}, and ChooseBest \cite{ICDE17LsmSSD} choose SSTs judiciously to reduce overlapped key range and reduce the data size of compactions. 
\LOGO{Luda} is orthogonal to these proposals.

Researches on GPUs in the storage community mainly focus on exploring the computational potential of GPUs and optimizing the data transfer path.

\textit{Utilize GPU for computation:}
Shredder \cite{FAST12Shredder} uses GPUs for data deduplication and incremental data processing.
Mega-KV \cite{VLDB15Mega-KV} adopts GPUs to accelerate the operations of in-memory key-value stores.
\LOGO{Luda} is the first to use GPUs for key value store compactions.

\textit{Optimize data path:}
Before NVIDIA GPUDirect Storage \cite{URL19GPUDirectStorage} and AMD SSG \cite{URL16AMDSSG}, directly moving data from storage devices to GPUs requires extra efforts, like GPUfs \cite{ASPLOS15GPUfs}.
DRAGON \cite{SC18DRAGON} directly maps NVM storage devices to GPU address space and enlarges the GPU address space.

\section{Conclusion}
\label{sec-conclusion}

In this paper, we develpo a GPU-accelerated compaction key value store \LOGO{Luda}.
\LOGO{Luda} utilizes the GPU computation power to boost compaction operations by subdividing SSTs to data independent blocks and processing them in parallel.
\LOGO{Luda} minimizes the overhead of data movements to accommodate different memory access characteristics of the GPU as compared to CPU.
Finally, we conduct evaluations of our prototype on a commodity GPU with CUDA, and also demonstrate the robustness and improved performance of \LOGO{Luda} compared to LevelDB and RocksDB under different levels of CPU overhead.


\bibliographystyle{IEEEtranS}
\bibliography{cluster20}

\begin{thebibliography}{10}
\providecommand{\url}[1]{#1}
\csname url@samestyle\endcsname
\providecommand{\newblock}{\relax}
\providecommand{\bibinfo}[2]{#2}
\providecommand{\BIBentrySTDinterwordspacing}{\spaceskip=0pt\relax}
\providecommand{\BIBentryALTinterwordstretchfactor}{4}
\providecommand{\BIBentryALTinterwordspacing}{\spaceskip=\fontdimen2\font plus
\BIBentryALTinterwordstretchfactor\fontdimen3\font minus
  \fontdimen4\font\relax}
\providecommand{\BIBforeignlanguage}[2]{{%
\expandafter\ifx\csname l@#1\endcsname\relax
\typeout{** WARNING: IEEEtranS.bst: No hyphenation pattern has been}%
\typeout{** loaded for the language `#1'. Using the pattern for}%
\typeout{** the default language instead.}%
\else
\language=\csname l@#1\endcsname
\fi
#2}}
\providecommand{\BIBdecl}{\relax}
\BIBdecl

\bibitem{URL16AMDSSG}
``{AMD Radeon Pro SSG Set to Transform Workstation PC Architecture, and to
  Shatter Real-Time Visual Computing Barriers.}''
  \url{http://www.amd.com/en-us/press-releases/Pages/amd-radeon-pro-2016jul25.aspx}.

\bibitem{URL19CUDA}
``{CUDA Toolkit Documentation},''
  \url{https://docs.nvidia.com/cuda/archive/10.1/}.

\bibitem{URL19GPUDirectStorage}
``{GPUDirect Storage: A Direct Path Between Storage and GPU Memory},''
  \url{https://devblogs.nvidia.com/gpudirect-storage/}.

\bibitem{URL19LevelDB}
``{LevelDB},'' \url{https://github.com/google/leveldb}.

\bibitem{URL19RocksDB}
``{RocksDB},'' \url{https://github.com/facebook/RocksDB}.

\bibitem{URL19stress-ng}
``{stress-ng},'' \url{https://kernel.ubuntu.com/git/cking/stress-ng.git/}.

\bibitem{VLDB15Compaction}
M.~Y. Ahmad and B.~Kemme, ``{Compaction management in distributed key-value
  datastores},'' in \emph{41st International Conference on Very Large Data
  Bases (VLDB'15)}, Hohala Coast, HW, August 2015.

\bibitem{ATC17TRIAD}
O.~Balmau, D.~Didona, R.~Guerraoui, W.~Zwaenepoel, H.~Yuan, A.~Arora, K.~Gupta,
  and P.~Konka, ``{{TRIAD}: Creating Synergies Between Memory, Disk and Log in
  Log Structured Key-Value Stores},'' in \emph{2017 USENIX Annual Technical
  Conference (USENIX ATC'17)}, Santa Clara, CA, July 2017.

\bibitem{ATC19SILK}
O.~Balmau, F.~Dinu, W.~Zwaenepoel, K.~Gupta, R.~Chandhiramoorthi, and
  D.~Didona, ``{SILK: Preventing Latency Spikes in Log-Structured Merge
  Key-Value Stores},'' in \emph{2019 USENIX Annual Technical Conference
  (ATC'19)}, Renton, WA, July 2019.

\bibitem{FAST12Shredder}
{Bhatotia, Pramod and Rodrigues, Rodrigo, and Verma, Akshat}, ``{Shredder:
  GPU-Accelerated Incremental Storage and Computation},'' in \emph{10th USENIX
  Conference on File and Storage Technologies (FAST'12)}, Santa Clara, CA,
  February 2012.

\bibitem{ICDE19LDCCompaction}
Y.~Chai, Y.~Chai, X.~Wang, H.~Wei, N.~Bao, and Y.~Liang, ``{LDC: A Lower-Level
  Driven Compaction Method to Optimize SSD-Oriented Key-Value Stores},'' in
  \emph{IEEE 35th International Conference on Data Engineering (ICDE'19)},
  Macao, China, April 2019.

\bibitem{ATC18HashKV}
H.~H.~W. Chan, Y.~Li, P.~P.~C. Lee, and Y.~Xu, ``{HashKV: Enabling Efficient
  Updates in {KV} Storage via Hashing},'' in \emph{2017 USENIX Annual Technical
  Conference (USENIX ATC'17)}, Boston, MA, July 2018.

\bibitem{SoCC10BechmarkYCSB}
B.~F. Cooper, A.~Silberstein, E.~Tam, R.~Ramakrishnan, and R.~Sears,
  ``{Benchmarking cloud serving systems with YCSB},'' in \emph{1st ACM
  symposium on Cloud computing (SoCC'10)}, Indianapolis, IN, June 2010.

\bibitem{SOSP07Dynamo}
{DeCandia, Giuseppe and Hastorun, Deniz and Jampani, Madan and Kakulapati,
  Gunavardhan and Lakshman, Avinash and Pilchin, Alex and Sivasubramanian,
  Swaminathan and Vosshall, Peter and Vogels, Werner}, ``{Dynamo: amazon's
  highly available key-value store},'' in \emph{21th ACM SIGOPS symposium on
  Operating systems principles (SOSP'07)}, Stevenson, WA, USA, October 2007.

\bibitem{SIGMOD19X-Engine}
G.~Huang, X.~Cheng, J.~Wang, Y.~Wang, D.~He, T.~Zhang, F.~Li, S.~Wang, W.~Cao,
  and Q.~Li, ``{X-Engine: An Optimized Storage Engine for Large-scale
  E-commerce Transaction Processing},'' in \emph{38th ACM SIGMOD International
  Conference on Management of Data (SIGMOD'19)}, Amsterdam, Netherlands, June
  2019.

\bibitem{MICRO17Summarizer}
G.~Koo, K.~K. Matam, T.~I, H.~V. K.~G. Narra, J.~Li, H.-W. Tseng, S.~Swanson,
  and M.~Annavaram, ``{Summarizer: trading communication with computing near
  storage},'' in \emph{50th Annual IEEE/ACM International Symposium on
  Microarchitecture (MICRO'17)}, Cambridge, Massachusetts, October 2017.

\bibitem{SOSP19KVell}
B.~Lepers, O.~Balmau, K.~Gupta, and W.~Zwaenepoel, ``{KVell: the Design and
  Implementation of a Fast Persistent Key-Value Store},'' in \emph{ACM
  Symposium on Operating Systems Principles (SOSP'19)}, Huntsville, Ontario,
  Canada, October 2019.

\bibitem{FAST16WiscKey}
L.~Lu, T.~S. Pillai, A.~C. Arpaci-Dusseau, and R.~H. Arpaci-Dusseau,
  ``{WiscKey: Separating Keys from Values in SSD-Conscious Storage},'' in
  \emph{14th USENIX Conference on File and Storage Technologies (FAST'16)},
  Santa Clara, CA, February 2016.

\bibitem{arXiv19LsmStability}
C.~Luo and M.~J. Carey, ``{On Performance Stability in LSM-based Storage
  Systems},'' \emph{arXiv}, vol. abs/1906.09667, June 2019.

\bibitem{SC18DRAGON}
P.~Markthub, M.~E. Belviranli, S.~Lee, J.~S. Vetter, and S.~Matsuoka,
  ``{DRAGON: breaking GPU memory capacity limits with direct NVM access},'' in
  \emph{International Conference for High Performance Computing, Networking,
  Storage, and Analysis (SC'18)}, Dallas, TX, November 2018.

\bibitem{SoCC18SifrDB}
F.~Mei, Q.~Cao, H.~Jiang, and J.~Li, ``{SifrDB: Unified Solution for
  Write-Optimized Key-Value Stores in Large Datacenter},'' in \emph{ACM
  Symposium on Cloud Computing (SoCC'18)}, Carlsbad, PA, October 2018.

\bibitem{FAST12SFS}
{Min, Changwoo and Kim, Kangnyeon and Cho, Hyunjin and Lee, Sang-Won and Eom,
  Young Ik}, ``{SFS: Random Write Considered Harmful in Solid State Drives},''
  in \emph{10th USENIX Conference on File and Storage Technologies (FAST'12)},
  Santa Clara, CA, February 2012.

\bibitem{Springer17dCompaction}
F.~Pan, Y.~Yue, and J.~Xiong, ``{dCompaction: Delayed Compaction for the
  LSM-Tree},'' \emph{interwordnational Journal of Parallel Programming},
  vol.~45, no.~6, pp. 1310--1325, 2017.

\bibitem{SOSP17PebblesDB}
P.~Raju, R.~Kadekodi, V.~Chidambaram, and I.~Abraham, ``Pebblesdb: Building
  key-value stores using fragmented log-structured merge trees,'' in \emph{ACM
  Symposium on Operating Systems Principles (SOSP'17)}, Shanghai, China,
  October 2017.

\bibitem{FAST14LogDRAM}
S.~M. Rumble, A.~Kejriwal, and J.~Ousterhout, ``{Log-Structured Memory for
  DRAM-based Storage},'' in \emph{12th USENIX Conference on File and Storage
  Technologies (FAST'14)}, Santa Clara, CA, February 2014.

\bibitem{ASPLOS15GPUfs}
M.~Silberstein, B.~Ford, I.~Keidar, and E.~Witchel, ``{GPUfs: Integrating a
  File System with GPUs},'' in \emph{18th International Conference on
  Architectural Support for Programming Languages and Operating Systems
  (ASPLOS'13)}, Houston, Texas, March 2013.

\bibitem{ICPP19TStore}
H.~Sun, W.~Liu, J.~Huang, S.~Fu, Z.~Qiao, and W.~Shi, ``{Near-Data
  Processing-Enabled and Time-Aware Compaction Optimization for LSM-tree-based
  Key-Value Stores},'' in \emph{48th International Conference on Parallel
  Processing (ICPP'19)}, Kyoto, Japan, August 2019.

\bibitem{arXiv18Co-KV}
H.~Sun, W.~Liu, J.~Huang, and W.~Shi, ``{Co-KV: A Collaborative Key-Value Store
  Using Near-Data Processing to Improve Compaction forthe LSM-tree},''
  \emph{arXiv}, vol. 1807.04151, July 2018.

\bibitem{FAST20FPGA-LSM}
{Teng Zhang and Jianying Wang and Xuntao Cheng and Hao Xu and Nanlong Yu and
  Gui Huang and Tieying Zhang and Dengcheng He and Feifei Li and Wei Cao and
  Zhongdong Huang and Jianling Sun}, ``{FPGA-Accelerated Compactions for
  LSM-based Key-Value Store},'' in \emph{18th USENIX Conference on File and
  Storage Technologies (USENIX FAST'20)}, Santa Clara, CA, Feburary 2020.

\bibitem{ICDE17LsmSSD}
R.~Thonangi and J.~Yang, ``{On Log-Structured Merge for Solid-State Drives},''
  in \emph{33rd IEEE International Conference on Data Engineering (IEEE
  ICDE'17)}, San Diego, CA, April 2017.

\bibitem{DATE18KVSSD}
S.-M. Wu, K.-H. Lin, and L.-P. Chang, ``{KVSSD: Close integration of LSM trees
  and flash translation layer for write-efficient KV store},'' in \emph{2018
  Design, Automation \& Test in Europe Conference \& Exhibition (DATE'18)},
  Dresden, Germany, March 2018.

\bibitem{MSST17Wan}
T.~Yao, J.~Wan, P.~Huang, X.~He, Q.~Gui, F.~Wu, and C.~Xie, ``{A light-weight
  compaction tree to reduce I/O amplification toward efficient key-value
  stores},'' in \emph{33rd International Conference on Massive Storage Systems
  and Technology (MSST'2017)}, Santa Clara, May 2017.

\bibitem{VLDB15Mega-KV}
K.~Zhang, K.~Wang, Y.~Yuan, L.~Guo, R.~Lee, and X.~Zhang, ``Mega-kv: a case for
  gpus to maximize the throughput of in-memory key-value stores,''
  \emph{Proceedings of the VLDB Endowment (VLDB'15)}, vol.~8, no.~11, pp.
  1226--1237, 2015.

\bibitem{IPDPS14LSM}
Z.~Zhang, Y.~Yue, B.~He, J.~Xiong, M.~Chen, L.~Zhang, and N.~Sun, ``{Pipelined
  Compaction for the LSM-tree},'' in \emph{Proceedings of the 28th IEEE
  International Parallel and Distributed Processing Symposium (IPDPS'14)},
  Phoenix, AZ, May 2014.

\end{thebibliography}

\begin{thebibliography}{00}
\bibitem{b1} G. Eason, B. Noble, and I. N. Sneddon, ``On certain integrals of Lipschitz-Hankel type involving products of Bessel functions,'' Phil. Trans. Roy. Soc. London, vol. A247, pp. 529--551, April 1955.
\bibitem{b2} J. Clerk Maxwell, A Treatise on Electricity and Magnetism, 3rd ed., vol. 2. Oxford: Clarendon, 1892, pp.68--73.
\bibitem{b3} I. S. Jacobs and C. P. Bean, ``Fine particles, thin films and exchange anisotropy,'' in Magnetism, vol. III, G. T. Rado and H. Suhl, Eds. New York: Academic, 1963, pp. 271--350.
\bibitem{b4} K. Elissa, ``Title of paper if known,'' unpublished.
\bibitem{b5} R. Nicole, ``Title of paper with only first word capitalized,'' J. Name Stand. Abbrev., in press.
\bibitem{b6} Y. Yorozu, M. Hirano, K. Oka, and Y. Tagawa, ``Electron spectroscopy studies on magneto-optical media and plastic substrate interface,'' IEEE Transl. J. Magn. Japan, vol. 2, pp. 740--741, August 1987 [Digests 9th Annual Conf. Magnetics Japan, p. 301, 1982].
\bibitem{b7} M. Young, The Technical Writer's Handbook. Mill Valley, CA: University Science, 1989.
\end{thebibliography}

%
%

\end{document}